\documentclass[12pt,twoside,a4paper,bibtex]{article}
\usepackage[dvips]{epsfig}
\newcommand{\widebar}[1]{%
   \mkern1.5mu\overline{\mkern-1.5mu#1\mkern-1.mu}\mkern1.mu}
\catcode`\@=11 

\newcommand{\tev}{\rm TeV}
\newcommand{\gev}{\rm GeV}
\newcommand{\mev}{\rm GeV}

\begin{document}
\thispagestyle{empty} 
\title{
\vskip-3cm
{\baselineskip14pt
\centerline{\normalsize DESY 01--031 \hfill ISSN 0418--9833}
\centerline{\normalsize hep--ph/0103091 \hfill} 
\centerline{\normalsize March 2001 \hfill}} 
\vskip1.5cm
Jet Photoproduction at THERA
\author{M.~Klasen
\vspace{2mm} \\
{\normalsize II. Institut f\"ur Theoretische Physik, Universit\"at Hamburg,}\\ 
\normalsize{Luruper Chaussee 149, 22761 Hamburg, Germany} \vspace{2mm}
\\ 
} }

\date{}
\maketitle
\begin{abstract}
\medskip
\noindent
We demonstrate that a future high-energy electron-proton collider like THERA
could largely extend the current HERA program in jet photoproduction of
testing QCD and determining the partonic structure of the proton and the
photon. Depending on the electron beam energy (250--500 GeV) and the collider
mode ($ep$ or $\gamma p$), the range in the hard transverse energy scale of
the jets could be increased by a factor of 2--3 and the reach in the momentum
fraction $x$ of the partons in the proton or photon by at
least one order of magnitude. It would thus become possible to check the
determinations of the gluon density in the proton obtained in deep--inelastic
scattering experiments, to measure the gluon density in the photon down to
low values of $x$, and to study the QCD dynamics in multi-jet events.
\end{abstract}


\section{Motivation}

In electron-proton collisions at the DESY HERA collider, the exchange of
almost real photons is responsible for the largest fraction of the scattering
events. In a subclass of these photoproduction events, hard jets are produced
with large transverse energies. The presence of a hard scale then allows for
a comparison of the data with predictions based on perturbative QCD. \\

Measurements of inclusive jets, dijets, and three jets have been performed by
the H1 and ZEUS collaborations at HERA over the last nine years and were found
to be in qualitatively good agreement with these predictions. They can also be
used to extract the free parameters of the theory like the strong coupling
constant or the parton densities in the colliding proton and photon. In the
proton case, this information is complementary to determinations in
deep-inelastic electron- or neutrino scattering or lepton pair production in
hadronic collisions. The HERA deep-inelastic scattering data have been
particularly useful to pin down the previously unknown gluon density at low
values of the partonic momentum fraction $x$. However, it is important to
test this determination in a second independent process like photoproduction.
Information on the photonic parton densities is still very limited: Only the
quark distributions have been constrained in deep-inelastic electron-photon
scattering, and only at large $x$. Little is known about the gluon density in
the photon. \\

The determination of the parton densities in the proton and photon is thus
an important research goal in the jet photoproduction experiments at HERA and
also in photon-photon scattering at LEP2. Unfortunately, the experiments have
so far been limited to transverse jet energies which may be
too low to suppress the soft underlying event coming from the proton or photon
remnants or the non-perturbative effects affiliated with hadronization.
Furthermore, they are kinematically limited to relatively large values of
$x$. \\

It is the aim of this paper to demonstrate that both of these restrictions can
be overcome if the electron energy is raised and/or the exchanged photons are
produced by laser backscattering. This may be possible at a facility where a
high-energy electron beam from a
future linear electron accelerator like TESLA is collided with a high-energetic
proton beam like the one available at HERA. At such a `THERA' collider it will
thus be possible to reach smaller values of $x$ and larger hard scales at the
same time.

\section{Dijet Cross Section}

For the proton beam we choose an energy of $E_p = 920\,\gev$, at which HERA
is currently operating. For the electron beam, we start with the current HERA
energy of $E_e = 27.5\,\gev$. If only one arm of the future electron
accelerator is used, $250$ and $400\,\gev$ can be reached in the first and
second stages of TESLA, respectively. If both arms are used, the electron
energy can be raised to as much as $500\,\gev$ already in the first stage.
Photoproduction events are selected by requiring that the electron scattering
angle is less than $1^\circ$ and that the photon momentum fraction lies within
the range $0.2 < y < 0.85$. For the different electron beam energies, this
maximum scattering angle corresponds to maximum photon virtualities of
$0.18$, $15$, $39$, and $61\,\gev^2$ at low $y$. The alternative approach of
choosing a constant maximum virtuality of $1\,\gev^2$ leads to unrealistically
small values of $\sim 0.1^\circ$ at TESLA energies. Finally, we investigate
the potential of a THERA $\gamma p$ collider where highly energetic real
photons are produced by backscattering laser light off a $250\,\gev$ electron
beam. \\

In leading order of perturbative QCD, two partons with equal transverse
energies are produced, corresponding to two hard jets. The dijet cross section
is then given by
\begin{equation}
  \frac{\mbox{d}^3\sigma}{\mbox{d}E_T^2\mbox{d}\eta_1\mbox{d}\eta_2}
  = \sum_{a,b} x_\gamma y f_{a/e}(x_\gamma y,\mu_f^2) x_p f_{b/p}(x_p,\mu_f^2)
  \frac{\mbox{d}\sigma}{\mbox{d}t}(ab \rightarrow p_1p_2).
\end{equation}
In next-to-leading
order there may also be a third, softer jet. We can then use the average
transverse energy $\widebar E_T$ and the average rapidity $\widebar \eta$ of
the dijet system as observables and allow the two jets to differ in transverse
energy by as much as $\Delta E_T < \widebar E_T / 2$. This choice, which allows
for a full cancellation of infrared singularities and avoids the sensitive
region of two equal minimal $E_T$ \cite{pl:b366:385},
has also been made in a recent H1 analysis \cite{epj:c1:97}. The rapidity
difference of the two jets $\Delta\eta$ is related to the center-of-mass
scattering angle $\cos (\theta^\ast) = \tanh (\Delta\eta)/2$.
While inclusive jet measurements yield higher statistics, only dijet analyses
allow for a reconstruction of
\begin{eqnarray}
 x_p^{\rm obs}       =  \frac{E_{T,1} e^{+\eta_1}+E_{T,2}
                           e^{+\eta_2}}{2E_p} & , &
 x_\gamma^{\rm obs}  =  \frac{E_{T,1} e^{-\eta_1}+E_{T,2}
                           e^{-\eta_2}}{2yE_e}
\end{eqnarray}
which, in leading order, match exactly the momentum fractions of the partons
in the proton $x_p$ and photon $x_\gamma$, but neglect the contribution of a
possible third jet. Jet photoproduction has been calculated in next-to-leading
order QCD using three different phase space slicing methods
\cite{zfp:c76:67,epjd:c1:1,pr:d56:4007,pr:d57:5555,epj:c17:413} and the
subtraction method \cite{np:b507:315}. The results were found to agree with
each other within a few percent \cite{epj:c17:413,hep-ph-9905348}.
In our next-to-leading order
calculation \cite{zfp:c76:67,epjd:c1:1}, jets are defined according to the
$k_T$ cluster algorithm with the parameter $R = 1$
\cite{np:b406:187,pr:d48:3160}. For the parton densities in the photon
and proton, we choose the next-to-leading order set of GRV \cite{pr:d46:1973}
and the latest CTEQ parameterization 5M \cite{epj:c12:375} with the
corresponding value of $\Lambda^{\rm nf = 5}_{\overline{\rm MS}} = 226\,\mev$.
The strong coupling constant $\alpha_s$ is evaluated at two loops and at the
scale $\mu=\mu_f=\max(E_{T,1},E_{T,2})$. The sensitivity of a THERA collider
to different photon parton densities has been analyzed in detail elsewhere
\cite{thera-wing}. \\

With higher electron beam energies, the HERA center-of-mass energy $\sqrt{S}$
of $318\,\gev$ can be increased to $959$, $1213$, or even $1357\,\gev$,
approaching the $2\,\tev$ regime of the Fermilab Tevatron in Run 2. At the
Tevatron, jets with transverse energies in excess of $50\,\gev$ are selected.
Since THERA would operate at roughly half the Tevatron center-of-mass energy,
we cut on $\widebar E_T > 25\,\gev$.

\section{Results}

\begin{figure}[h]
 \begin{center}
  \epsfig{file=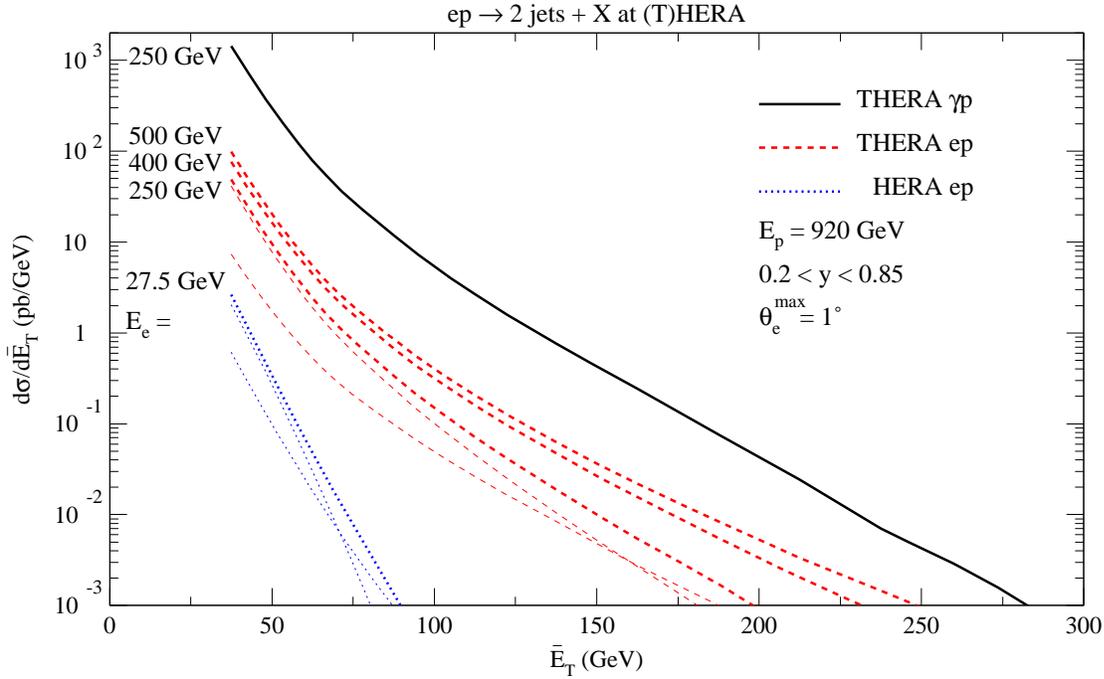,width=\textwidth}
 \end{center}
 \caption{Differential dijet photoproduction cross section as a function of
  the average transverse energy of the two jets $\widebar E_T$. The thin
  lines show the separate contributions from the resolved (direct) processes
  for HERA and THERA with $E_e=$250 \gev, which dominate at small (large)
  $\widebar E_T$.}
 \label{fig:klasen:1}
\end{figure}

Like hadronic jet cross sections, photoproduction jet cross sections drop
steeply in transverse energy. It is therefore interesting to study the
size of the dijet photoproduction cross section as a function of $\widebar
E_T$ as shown in Figure~\ref{fig:klasen:1}.
At $\widebar E_T = 40\,\gev$, THERA with electron beam energies of $250$ to
$500\,\gev$ produces cross sections which are larger than the HERA cross
section by about a factor of 20-40. At a THERA photon collider, the cross
section is even larger by a factor of 500. The larger cross sections result,
of course,
in a much extended range in $\widebar E_T$. With an expected luminosity of 100
pb$^{-1}$/year, the range can be extended from $75\,\gev$ at HERA to
$150\,\gev$ at
THERA $ep$ or $225\,\gev$ at THERA $\gamma p$, {\it i.e.} by a factor of
2-3. \\

\begin{figure}[h]
 \begin{center}
  \epsfig{file=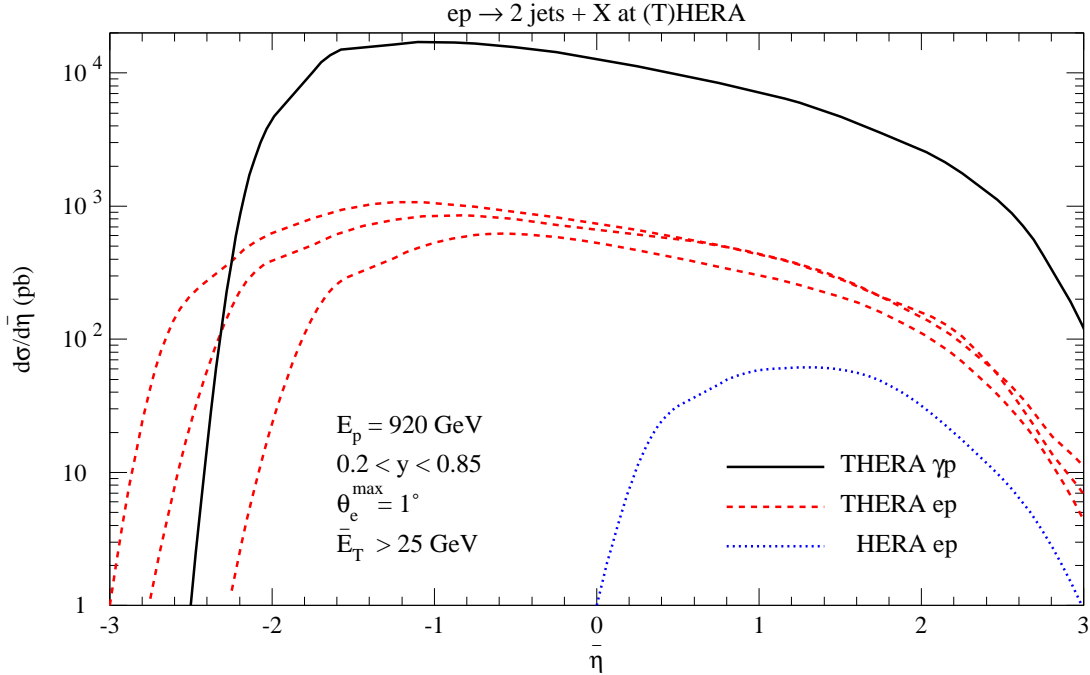,width=\textwidth}
 \end{center}
 \caption{Differential dijet photoproduction cross section as a function of
  the average rapidity of the two jets $\widebar \eta$. For THERA $ep$ we show
  results with three different electron beam energies $E_e=$250, 400, and
  500 \gev.}
 \label{fig:klasen:2}
\end{figure}

Of course, these jets need not only be produced but also be measured in a
detector. An important question in this context is the required coverage in
rapidity. In Figure~\ref{fig:klasen:2} we therefore show the average rapidity
distribution of the produced dijet system with $\widebar E_T > 25\,\gev$.
Jets at HERA are produced mostly in the proton (forward) direction $0 <
\widebar \eta < 3$, which has lead to the characteristic asymmetric designs of
the H1 and ZEUS detectors. In contrast, jets at THERA will be produced
centrally in the range $-3 < \widebar \eta < 3$, requiring a more symmetric
detector design, closer to the design used at hadron colliders.
Therefore, if the H1 and ZEUS detectors are to be used, some modifications
will be necessary. However, an upgrade of the electron beam energy from 250
to $500\,\gev$ will probably not necessitate additional changes, since the
rapidity range is then only slightly extended. \\

The rapidity difference of the two jets or, equivalently, the cosine of the
center-of-mass
scattering angle $\cos(\theta^\ast)$ is related to the $2\rightarrow 2$
Mandelstam variables of the underlying partonic subprocesses by
\begin{eqnarray}
 t = -\frac{1}{2}s(1-\cos\theta^\ast) & , &
 u = -\frac{1}{2}s(1+\cos\theta^\ast),
\end{eqnarray}
where $s=(p_a+p_b)^2=x_\gamma y x_p S$ is the partonic center-of-mass energy
squared.
Most of the resolved (parton-parton) scattering processes are characterized
by the exchange of a massless vector boson in the $t$-channel
\begin{equation}
 \overline{|{\cal M}|^2} \propto t^{-2} = \left[ -\frac{1}{2} s (1 -
 \cos\theta^\ast) \right]^{-2},
\end{equation}
whereas the direct processes proceed through a massless fermion
exchange in the $t$-channel with less singular behavior,
\begin{equation}
 \overline{|{\cal M}|^2} \propto t^{-1} = 
   \left[ -\frac{1}{2} s(1-\cos\theta^\ast)\right]^{-1}
\end{equation}
\begin{figure}[h]
 \begin{center}
  \epsfig{file=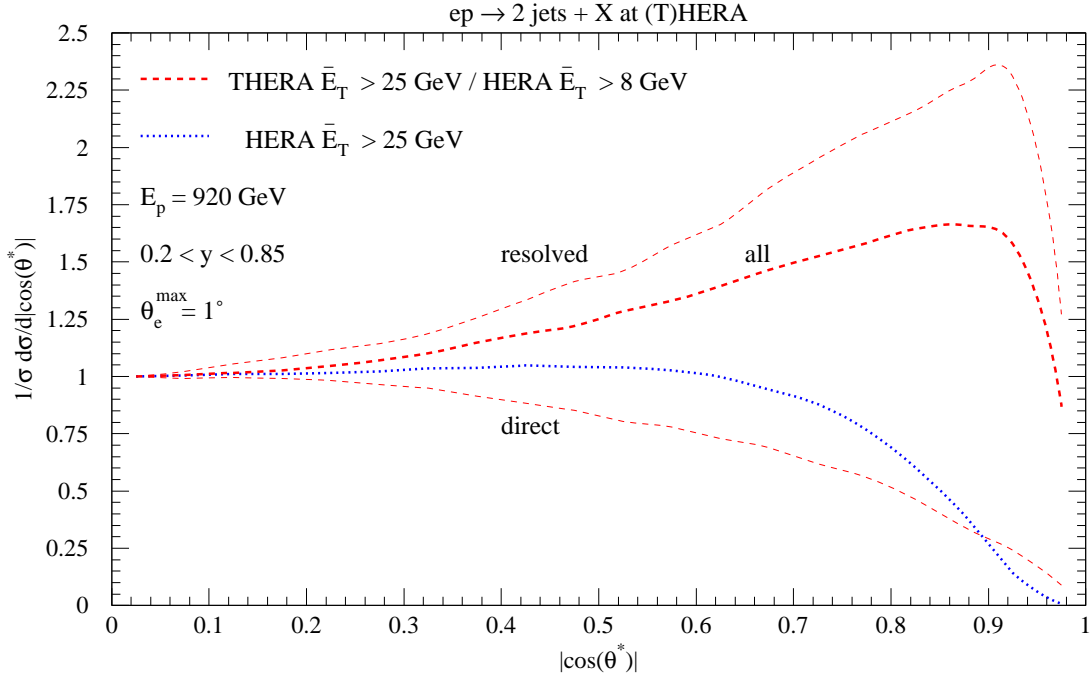,width=\textwidth}
 \end{center}
 \caption{Differential dijet photoproduction cross section as a function of
  the cosine of the center-of-mass scattering angle of the two jets
  $|\cos (\theta^\ast)|$. All curves have been normalized at $|\cos
  (\theta^\ast)|=0$.}
 \label{fig:klasen:3}
\end{figure}
or through $s$-channel contributions without any singular behavior.
In Figure~\ref{fig:klasen:3} we show the normalized dijet cross section
as a function of $|\cos(\theta^\ast)|$.
At HERA, the rather high cut on $\widebar E_T > 25\,\gev$ results in a
$|\cos(\theta^\ast)|$ distribution which is mostly dominated by phase space.
At THERA, the center-of-mass energies are larger: Phase space restrictions
are unimportant, and the normalized
distribution no longer depends on the electron beam energy or the collider
mode ($ep$ or $\gamma p$), if the $\widebar E_T$ cut is kept fixed.
Therefore the expected
singular behavior can now clearly be seen. The same distribution is valid if
the HERA cut is scaled down to $\widebar E_T > 318\,\gev /
959\,\gev \times 25\,\gev \simeq 8 \gev$, which is similar to
the cut $E_T > 6\,\gev$ used in a recent ZEUS analysis \cite{pl:b384:401}.
The ZEUS data were found to agree with next-to-leading order QCD
predictions \cite{pr:d56:4007,epj:c7:225}.
In Figure~\ref{fig:klasen:3} we also show
results for direct and resolved photoproduction separately (thin curves).
The resolved curve clearly shows the singular behavior in contrast to the
direct contribution, which contributes only a small fraction to the total
result (see also Figure~\ref{fig:klasen:1}).

For determinations of the partonic structure of protons and photons,
distributions in the observed partonic momentum fractions are of great value.
\begin{figure}[h]
 \begin{center}
  \epsfig{file=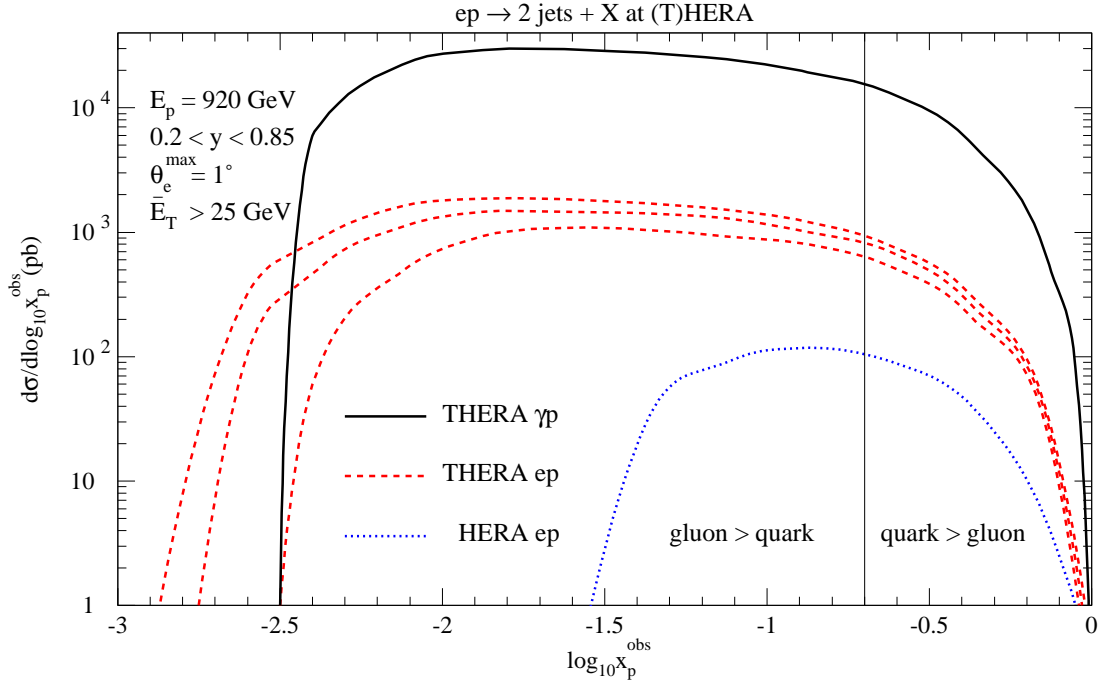,width=\textwidth}
 \end{center}
 \caption{Differential dijet photoproduction cross section as a function of
  the observed parton momentum fraction in the proton $x_p^{\rm obs}$. For
  THERA $ep$ we show results with three different electron beam energies
  $E_e=$250, 400, and 500 \gev.}
 \label{fig:klasen:4}
\end{figure}
Figure~\ref{fig:klasen:4} demonstrates that the range in $x_p^{\rm obs}$, in
which
the proton structure can be analyzed, is extended by at least one order of
magnitude from 0.03 at HERA to 0.003--0.001 at THERA. Since the gluon
dominates at values below 0.2, the low-$x$ gluon determinations in
deep-inelastic scattering could thus be tested in photoproduction for the
first time.

Similarly, the range in $x_\gamma^{\rm obs}$ would be extended by at least one
order of magnitude from 0.04 at HERA to 0.004--0.0025 at THERA.
\begin{figure}[h]
 \begin{center}
  \epsfig{file=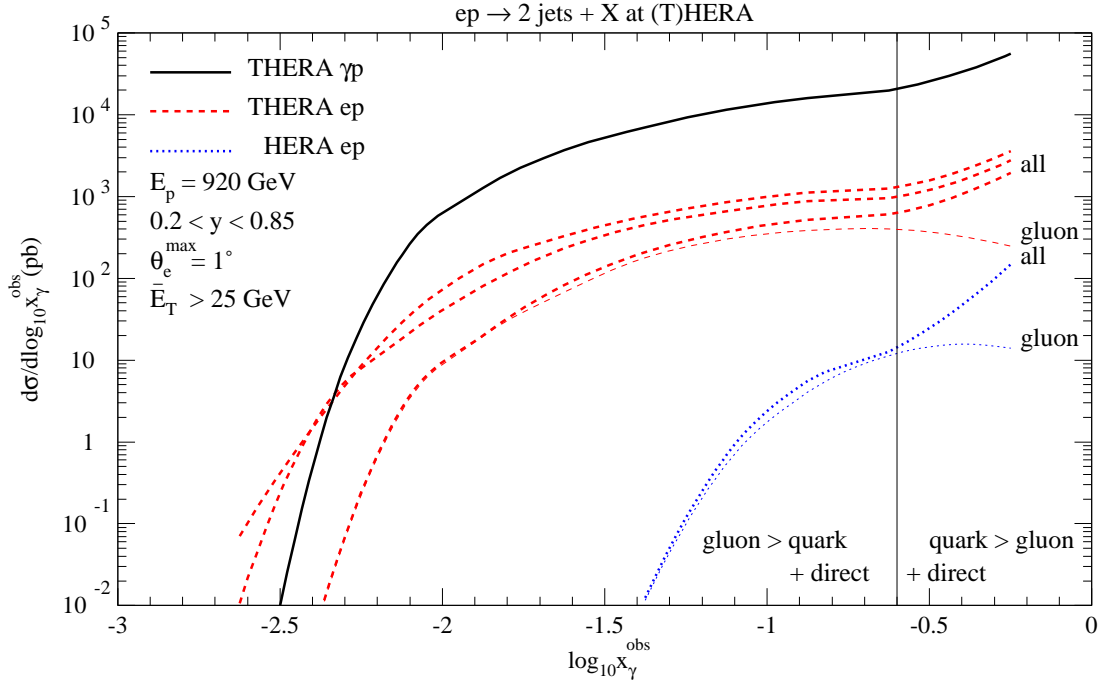,width=\textwidth}
 \end{center}
 \caption{Differential dijet photoproduction cross section as a function of
  the observed parton momentum fraction in the photon $x_\gamma^{obs}$.
  For THERA $ep$ we show results with three different electron beam energies
  $E_e=$250, 400, and 500 \gev.}
 \label{fig:klasen:5}
\end{figure}
This can be seen in Figure~\ref{fig:klasen:5}. In the photon case, the gluon
dominates below 0.25, so that HERA should already have the potential to
constrain the gluon in the photon with photoproduced high-$\widebar E_T$ jets
in the region 0.04--0.25. THERA could, however, constrain the gluon down to
much lower values of $x_\gamma^{\rm obs}$. THERA $ep$ cross sections with an
electron beam energy of $250\,\gev$ and different next-to-leading order
parameterizations for the gluon density in the photon differ by 30--50\% for
$E_T > 29$ and $14\,\gev$ \cite{thera-wing}.

\section{Conclusion}
Collisions of high-energy electrons from a future linear accelerator like
TESLA with an existing proton beam in a `THERA' $ep$ machine offer great
opportunities for jet photoproduction: They would naturally build on the
current
HERA program of testing QCD and determining the partonic structure of the
proton and the photon. The range in the hard scale $\widebar E_T$ could be
extended by a factor of 2--3, which would reduce complications from the soft
underlying event and hadronization. At the same time the range in the
partonic momentum fractions $x_p$ and $x_\gamma$ could be extended by at least
one order of magnitude. Determinations of the gluon in the proton at low $x_p$
in deep inelastic scattering could then be tested, and the gluonic structure of
the photon could be determined for the first time.
Furthermore, event rates with three or more observed jets in the final
state are expected to be larger at higher energies and
could be compared to then-available higher order  predictions to study the
multi-particle QCD dynamics.
The physics program of any future linear collider would thus greatly benefit
from these additional opportunities.

\section*{Acknowledgments}
The author thanks G.\ Kramer and H.\ Spiesberger for their encouragement and
useful discussions and the Deutsche Forschungsgemeinschaft and the European
Commission for financial support through Grants KL~1266/1-1 and
ERBFMRX-CT98-0194.

\end{document}